\documentclass[10pt]{wlscirep}
\usepackage{geometry}
\usepackage[utf8]{inputenc}
\usepackage[T1]{fontenc}

\usepackage{caption}
\usepackage{enumitem}
\usepackage{amsmath}
\usepackage{colortbl}
\usepackage{xcolor} 
\usepackage[framemethod=TikZ]{mdframed}
\usepackage[explicit]{titlesec}
\usepackage{tabularx}
\usepackage{calc} 
\definecolor{titlegray}{RGB}{167,169,172}

\definecolor{mybackgroundcolor}{RGB}{240,240,240} 
\newlength{\msectionboxwidth}
\titleclass{\msection}{straight}[\section]

\titleformat{\msection}[block]
  {\normalfont\bfseries\color{black}\centering}
  {}
  {0em}
  {\fcolorbox{black}{mybackgroundcolor}{\parbox{\msectionboxwidth}{\centering #1}}}
\titlespacing*{\msection}
  {0pt}
  {3.25ex plus 1ex minus .2ex}
  {1.5ex plus .2ex}

\title{LLM-Twin: Mini-Giant Model-driven Beyond 5G Digital Twin Networking Framework with Semantic Secure Communication and Computation}

\author[1]{Yang Hong}
\author[1,*]{Jun Wu}
\author[2]{Rosario Morello}
\affil[1]{Graduate School of Information, Production and System, Waseda University, Fukuoka, 8080135, Japan.}
\affil[2]{Department of Information Engineering, University "Mediterranea" of Reggio Calabria, Via Graziella, 89122, Reggio Calabria, Italy.}

\affil[*]{corresponding: jun.wu@ieee.org}

\begin{abstract}
Beyond 5G networks provide solutions for next-generation communications, especially digital twins networks (DTNs) have gained increasing popularity for bridging physical space and digital space. However, current DTNs networking frameworks pose a number of challenges especially when applied in scenarios that require high communication efficiency and multimodal data processing. First, current DTNs frameworks are unavoidable regarding high resource consumption and communication congestion because of original bit-level communication and high-frequency computation, especially distributed learning-based DTNs. Second, current machine learning models for DTNs are domain-specific (e.g. E-health), making it difficult to handle DT scenarios with multimodal data processing requirements. Last but not least, current security schemes for DTNs, such as blockchain, introduce additional overheads that impair the efficiency of DTNs. To address the above challenges, we propose a large language model (LLM) empowered DTNs networking framework, LLM-Twin. First, we design the mini-giant model collaboration scheme to achieve efficient deployment of LLM in DTNs, since LLM are naturally conducive to processing multimodal data. Then, we design a semantic-level high-efficiency, and secure communication model for DTNs. The feasibility of LLM-Twin is demonstrated by numerical experiments and case studies. To our knowledge, this is the first to propose LLM-based semantic-level digital twin networking framework.
\end{abstract}
\begin{document}

\flushbottom
\maketitle
%
%
\thispagestyle{empty}

\section*{Introduction}

The development of next-generation communication technologies (beyond 5G), as well as internet of everything (IoE) technologies, has paved the way for visions of smart cities \cite{wolf2022towards}, smart transportation \cite{yigit2023twinport}, smart homes, etc., but it has also resulted in a growing need for rapid processing of massive and diverse network data \cite{10148936}. As one of the most promising next-generation data-driven paradigms, digital twin networks (DTNs) \cite{luan2021paradigm} can efficiently process massive amounts of data and have demonstrated great application value in various areas such as real-time transportation safety assessment \cite{10107352}, intelligent city scheduling \cite{9267778}, industry remote control \cite{8477101}, etc. DTNs are able to create real-time digital replicas of the physical world for fine-grained modeling, analytics, and prediction, which reveals that DTNs are significantly reshaping the future network paradigm in terms of efficiency and intelligence \cite{9429703}.

The main characteristic of digital twins (DTs) is bi-directional communication \cite{9765576, yeon2023dtumos}, which is denoted as intra-twin communication and inter-twin communication in DTNs \cite{10090432}. Intra-twin communication denotes the communication between a physical entity and its digital twin, where the DT utilizes powerful cloud computing and machine learning algorithms to model and present insights on historical and real-time information collected from physical entities, which effectively addresses the limitation of insufficient local capabilities of physical entities. Unfortunately, intra-twin communication is very expensive on physical networks because of the large amount of communication and computational resources consumed to support real-time applications \cite{10012285}. In addition, constructing DTs by uploading detailed physical entity data to servers via intra-twin communication also leads to huge data security concerns \cite{luan2021paradigm}. Inter-twin communication denotes information sharing and communication among each DT in DTNs, which enables the physical entity to gain a larger perceptual domain and provides it with global decision-making \cite{9852383}. Furthermore, inter-twin communication obtains information in DTNs through data access, virtual links, etc., which breaks the limitation of physical links and mainly relies on the computational power of the cloud server to model the data transmission behavior. However, due to the massive number of IoT devices and extremely diverse data types, it is also hard to model efficient data sharing for inter-twin communication. Therefore, realizing more efficient communication, intelligent computation, and more secure data processing capabilities in DTNs are still open issues at present \cite{9854866}.

Lu et al \cite{9164912} proposed digital twin edge networks (DTENs), which bridge the gap between physical edge networks and DT cloud servers by introducing distributed federated learning (FL) into DTNs. Specifically, distributed edge physical entities participate in the model computation and aggregate the parameters in the DT server, liberating the computational burden of the DT server. However, since FL requires multiple rounds of weight updating and aggregation introduces additional communication overhead \cite{10100949}. Therefore, asynchronous FL \cite{9145588}, low-latency FL \cite{9170905}, etc. investigate more efficient communication and computation of DTNs. In addition, reinforcement learning \cite{10142965}, and graph neural networks \cite{10253478} have been used to further reduce resource consumption and improve the reliability of DTNs. Furthermore, to address data privacy and security issues in DTNs, \cite{9606227, 10234720} joint FL and blockchain techniques to protect local model updates and global model updates against data tampering and privacy leakage.

In general, the existing works have explored efficient and intelligent communication computing schemes for DTNs and also considered data security issues, but limited by the traditional networking framework of DTNs, the following three challenges still exist:

\begin{enumerate}
\item Current works have not fundamentally changed the traditional DTNs networking framework of bit-level real-time communication and high-frequency computation, especially the high-frequency parameter communication and model updating of distributed FL, which is unavoidable regarding high physical resource consumption and communication congestion. In addition, current works have not considered the unifying design of communication and computation, which are executed independently and on a queue-by-queue basis. It results in communication delays and computation overheads overlapping each other, e.g. when a single node's computational power is weak, it will result in communication delays for the server to receive the parameter updates, which will ultimately delay the global model aggregation.
\item Traditional DTNs in which the machine learning models are domain-specific make it hard to handle DT scenarios with hybrid data types and with multimodal data processing requirements, which significantly limits the application scenarios and scalability of DTNs.
\item Traditional security strategies for DTNs, such as blockchain, differential privacy federated learning, and other schemes bring additional communication and computation overheads that cannot be balanced with efficiency, which greatly affects the network performance and quality of service of DTNs.
\end{enumerate}

\begin{figure}[ht]
\centering
\includegraphics[width=\linewidth]{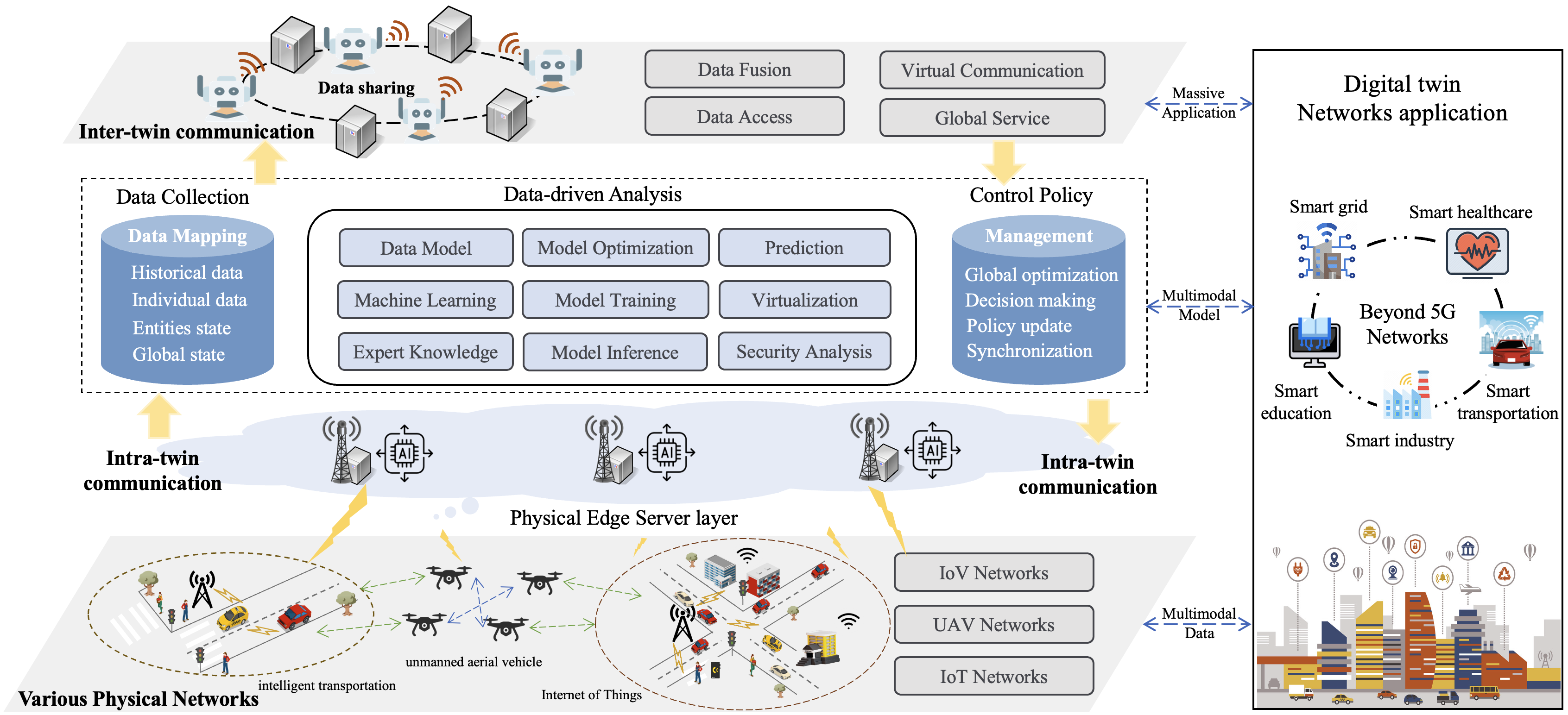}
\caption{Beyond 5G digital twin networks.}
\label{fig1}
\end{figure}

To address the above challenges, we propose introducing large language models (LLMs) in DTNs. LLMs are transformer language models containing hundreds of billions of parameters, such as ChatGPT \cite{10113601}, which demonstrate a strong ability to communicate with humans, the ability to handle mathematical problems, and a rich knowledge base. On top of that, as powerful generative models, LLMs demonstrate advanced capabilities in text-to-code \cite{zhong2023chatgpt}, text-to-image \cite{lai2023minidalle3}, text-to-speech \cite{rubenstein2023audiopalm}, and text-to-video \cite{hong2023large}. The excellent properties of LLMs provide advanced insights into the new generation of network communications, especially semantic communications \cite{raha2023generative}. Benefiting from the powerful multimodal generation capability of LLMs, the sender can compress the message $M$, which consumes a lot of communication resources, into $C$, and utilize LLMs to generate and reconstruct $M$ at the receiver, which can greatly reduce the overhead from bit-level communication in the traditional network. However, in contrast, the training, inference, and fine-tuning of LLMs imply a huge computation overhead, while offloading LLMs to edge devices also incurs a huge communication overhead. In particular, computation and communication resources are limited in IoT systems \cite{9854866}, thus this becomes a significant bottleneck for introducing generative LLMs in DTNs.

In this paper, we first introduce LLM-driven digital twin networking frameworks, LLM-Twin, which address the resource shortcomings of deploying LLMs in DTNs by designing a mini-giant model collaborative scheme. Second, we reshape the architectures of intra-twin communication and inter-twin communication to realize semantic-level DT communication and computation. Finally, we propose a strategy to strengthen the data security of LLM-Twin, which can realize the hiding of sensitive information in LLM. The main contributions of this paper are summarized as follows:

\begin{enumerate}
\item We propose the mini-giant LLMs collaboration scheme to solve the resource-limited problem of LLMs deployment in DTNs, which involves edge data fine-tuning and instruction prompts to achieve generalized intelligent DTNs, improving the ability of DTNs for multimodal data processing and multidomain generalized modeling.
\item We first propose digital twin semantic networks (DTSNs), which are semantic-level frameworks for efficient communication and computation in DTNs. Driven by LLM we reshape the traditional architectures of intra-twin communication and inter-twin communication, including static knowledge base synchronization and dynamic semantic-level information sharing, which realizes the unification of efficient communication and computation.
\item To further guarantee the security of the proposed LLM-Twin, we design an LLM data security strategy. It guarantees the privacy of sensitive information by reinforcing the reversal curse of LLM, which ensures that the sensitive information in the original input cannot be obtained in reverse from the service output of LLM-Twin. Finally, we give the security analysis and proof for LLM-Twin, while theoretical analysis and experiments show that our proposed networking framework significantly improves the communication computation efficiency of traditional DTNs.
\end{enumerate}

The rest of the article is structured as follows: first, the framework of the LLM-Twin is introduced, then the efficiency and security model for LLM-Twin is presented. Finally, the proposal is evaluated and this article is concluded.

\section*{Framework of LLM-Twin with efficiency communication and data security}

In this section, the proposed LLM-Twin framework will be presented in detail, further explaining how LLM-Twin addresses the existing challenges and makes contributions. The first subsection demonstrates the digital twin semantic networks (DTSNs), as shown in Fig. \ref{fig2}, which introduces the mini-giants model collaboration scheme for deploying LLMs in resource-limited DTNs. Then, we present a semantic-level redesign of the traditional intra-twin communication and inter-twin communication, which demonstrates the improvement in communication efficiency brought by LLM-Twin. In addition, since both LLM and DTN are based on large data-driven, this also poses a huge data security concern. Therefore, we discuss the data security model of the LLM-Twin in the second subsection and propose a data security protection scheme that does not bring additional communication overhead.

\subsection*{Digital twin semantic networks}

\subsubsection*{Mini-giants model collaboration scheme}
Compared to the domain specialization of traditional machine learning models, LLMs have unparalleled natural advantages in DT scenarios with a large number of hybrid data types and multimodal data processing requirements ( as shown in Fig. \ref{fig1}). However, the huge demand for computational resources in LLM makes it hard to apply in DTNs that consist of a large number of edge nodes and IoT devices and emphasize real-time performance. Zhou et al\cite{zhou2023minigiants} propose small language models (LMs), including reducing model parameters and fine-tuning a small number of parameters, which can be trained and used on affordable resources, such as a single GPU. In particular, small LMs based on LoRA \cite{hu2021lora}, QLoRA \cite{dettmers2023qlora} fine-tuning can approach the performance of LLMs by training only a very small fraction of parameters (one in ten thousand), and can be easily switched between different downstream tasks. Therefore, we first design the small LM scheme with the edge fine-tuning in DTNs.

\begin{figure}[ht]
\centering
\includegraphics[width=\linewidth]{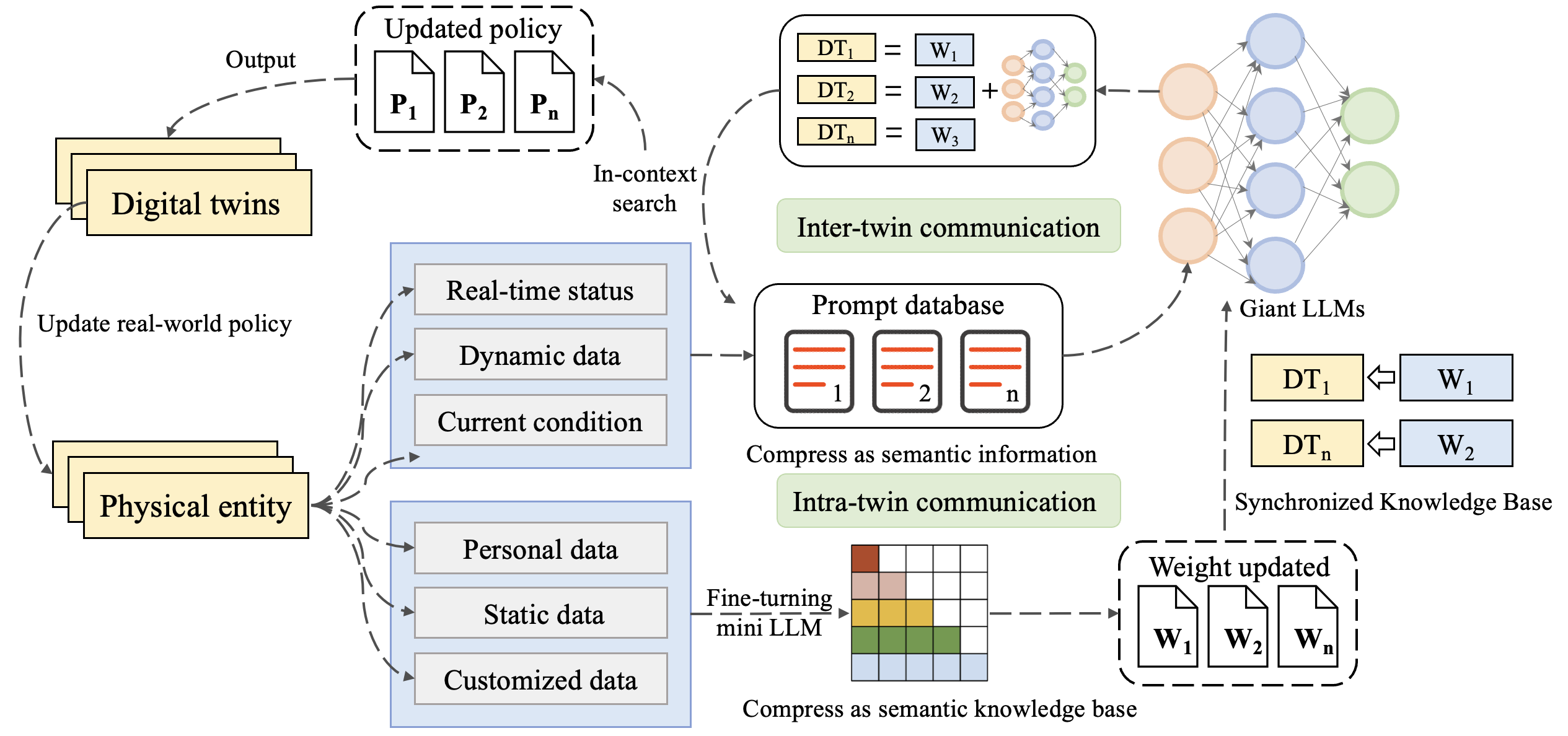}
\caption{Digital twin semantic network of LLM-Twin networking framework.}
\label{fig2}
\end{figure}

The core of this scheme is to address 1) state synchronization of DTs and physical entities; and 2) maintenance of DT decision models for entities, which are two important issues for DTNs. Firstly, pre-trained LLMs for specific scenarios are deployed in the DT cloud servers and distributed to the edge physical entities in the region. As shown in Fig. \ref{fig2}, the physical entity collects its own information to perform edge fine-tuning of the pre-trained LLM locally and obtains a small LM containing personal data and decision-making information with affordable computational cost. Subsequently, the small LM is loaded into the LLM by means of parameter updating to realize model updating and maintenance of the DT. In addition, for real-time physical entity state synchronization, we designed the instruction prompt scheme. Physical entities collect dynamic data and state information in real time and encode this information as prompts which are uploaded to the prompt database of the DT server.

When DT performs data analysis and decision-making, it conducts the following operations: 1) loads the small LM of the corresponding physical entity; 2) searches the prompt database to obtain the prompt of the corresponding physical entity, and adds it to the current analysis prompts as a context; and 3) sends the aggregated prompts to the LLM, then the RESPONSE is the final decision-making information, which is based on the current data model (the small LM of the physical entity) and the current state information (the physical entity's uploaded prompts). Overall, the proposed scheme addresses the deployment of LLM in DTNs, which reduces the resource overhead of edge devices and cloud servers.

\subsubsection*{Intra-twin communication}

Based on the above scheme, we further proposed DTSN framework with regard to the core communication subsystems of DTN, intra-twin communication and inter-twin communication.
Intra-twin communication is responsible for enabling fine-grained mapping of physical entities and digital twins. Traditional bit-level communication schemes need to maintain real-time communication and modeling mapping of huge data volumes, which incurs a huge communication overhead. In order to reduce unnecessary information transmission, semantic communication can achieve optimization of communication information based on a shared semantic knowledge base (KB) \cite{raha2023generative}. In DTSN, the static data of physical entities and individual preference data are modeled locally as small LM by means of edge fine-tuning as shown in Fig. \ref{fig2}, which forms its semantic KB. The KB will be uploaded and stored in the DT server. After the DTSN is established, the dynamic information of the physical entities will be encoded as semantic information (prompt) and synchronized to the DT, and the LLM of the DT loading the KB can parse the semantic information and then perform analysis and decision-making. In other words, the KB of the physical entity, the dynamic information (prompt) of the physical entity, and the LLM used for loading constitute the DT of this physical entity.

Compared to traditional communication, DTSN only needs to establish real-time channels for transmitting the compressed semantic information and update the KBs periodically according to the actual situation of physical entities, which greatly reduces the communication overhead. 

\subsubsection*{Inter-twin communication}

Inter-twin communication is responsible for enabling information sharing between DTs, which can enable entities to obtain global information and a larger perceptual field without physical communication constraints. The overhead of traditional inter-twin communication is mainly in two aspects: 1) the need to model the communication between different DTs, especially for cross-domain data and multimodal data processing and conversion will bring additional overhead; 2) although the communication is through the virtual link, the nodes or processes between the data transmission, data computation, and decision-making generation will still bring additional communication overhead.

Therefore, we model inter-twin communication in DTSN as a process in which LLM selects the prompts of the communicating parties from the prompt database to add to the context and execute them. First, since the communication data of each DT is encoded locally into a semantic representation (prompt) and then uploaded to the prompt database, there is no need to do further conversion of each modal data in inter-twin communication. Secondly, LLM-Twin does not need to establish additional virtual links, it adds the prompts of the communicating parties directly to the context of executing instructions to start inference and obtain new decisions, which unifies the sharing of information, data computation, and global decision updating.

\subsection*{Data security model}

Benefiting from the semantic communication performance and efficient resource allocation of LLM-Twin, it significantly benefits the communication, analysis, and processing of massive data in various DTN scenarios. However, it also brings data security risks. Therefore, in this section, we discuss the data security models of LLM-Twin, which have been taken into account during the design, including the homomorphic model and the one-way security model that enables privacy-preserving.

\begin{figure}[ht]
\centering
\includegraphics[width=16cm]{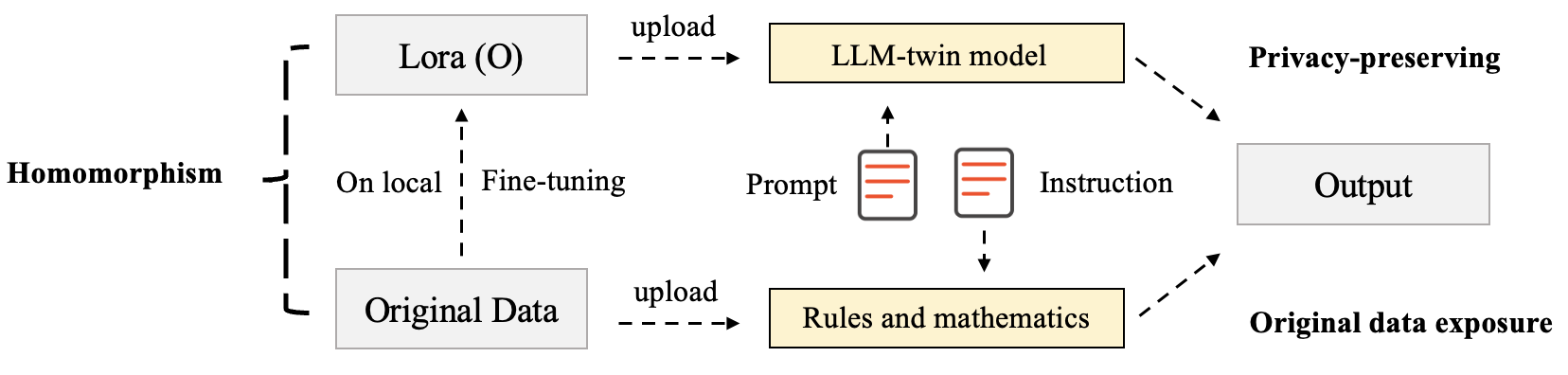}
\caption{Homomorphism model of LLM-Twin to protect original data.}
\label{fig3}
\end{figure}

\subsubsection{Homomorphism model}
The data communication and processing of LLM-Twin are homomorphic, as shown in Fig. \ref{fig3}. Homomorphism is manifested in that original data of physical entities can be explicitly uploaded into the rule-based and mathematical modeling DT for computation and obtaining specific results; similarly, the original data can be locally fine-tuned, e.g., by Lora, be converted into model weight information and then loaded into the LLM of the DT for execution and obtain the same results as those computed in the plaintext. Therefore, in the LLM-Twin framework, the private information of a physical entity appears only in the local computing environment of the physical entity and exists in non-plaintext form in both communication and computation. As a result, the homomorphism model protects the privacy of personal data while ensuring the consistency of computation results.

It is worth mentioning that in LLM-Twin we assume that the sensitive information is loaded into the DT's LLM by fine-tuning, as shown in Fig. \ref{fig2}. However, the real-time dynamic data of the physical entities encoded as prompt communication does not contain sensitive information as it is to be shared with other DTs.The protection of this type of information involves research such as privacy computing and is beyond the scope of this paper.

\begin{figure}[ht]
\centering
\includegraphics[width=16cm]{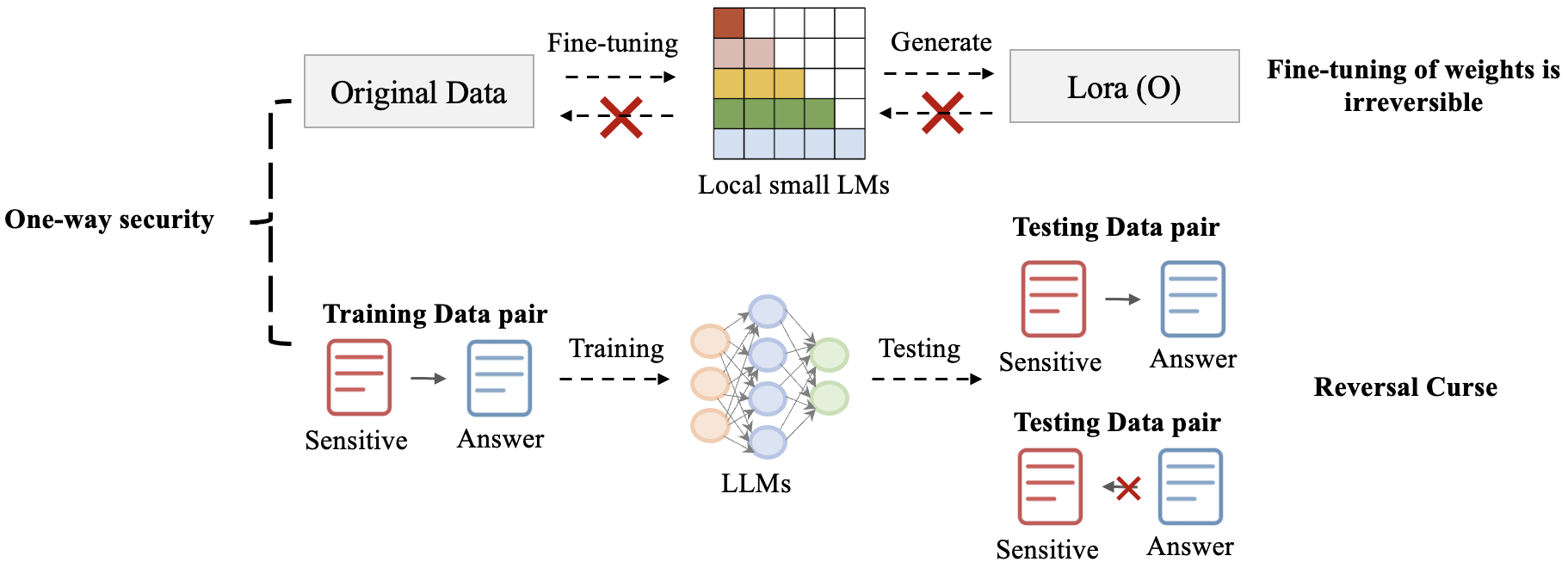}
\caption{One-way security model of LLM-Twin.}
\label{fig4}
\end{figure}

\subsubsection{One-way security model}

The one-way security of LLM-Twin is mainly manifested in two parts, as shown in Fig. \ref{fig4}. First, the fine-tuning of LLM is irreversible. The original data can be fine-tuned to train the small LM, but the small LM cannot restore the original data. Further, the local small LM generates the fine-tuned weight file $Lora$ which is uploaded to the DT side, but the $Lora$ file cannot directly restore the original data.

Second, there is a reversal curse in LLM \cite{berglund2023reversal}, this study reveals that LLM trained in the fixed pattern $<A\ is\ B>$ cannot answer $<B\ is\ ? >$. This research was intended to reveal the flaws in the logical capabilities of LLM, but we believe that reinforcing this property in LLM-Twin can help secure sensitive data. As previously described, sensitive information of physical entities in LLM-Twin, including personal data, personalization settings, etc., will be loaded into the LLM of DT in the form of edge fine-tuning to complete model construction and KB synchronization of DT. Although this information is loaded into the LLM in a non-explicit form, a malicious could deduce the original information through continuous interaction with the LLM. Therefore, we will design a training data format to strengthen the reversal curse by using only data forms such as $<Sensitive\ to\ Answer>$, which makes it impossible for a malicious to get the sensitive information by colliding a large number of $<Answer>$. As a result, LLM-Twin contains a homomorphic and one-way security design that incorporates data security without adding additional communication overhead.

\section*{Efficiency analysis and security models of LLM-Twin}

In this section, we present a detailed mathematical modeling of the computation and communication model of LLM-Twin presented in the previous section, as shown in Fig. \ref{fig2}, which demonstrates the advantages of the high efficiency of LLM-Twin. Specifically, we analyze the traditional FL-based networking framework of DTNs and our approach, which demonstrates the superiority of the proposed approach by comparing the time to complete one DTN modeling, $ T_{LLM-Twin} $ and $ T_{fl} $. Meanwhile, we give a macro security analysis of the whole protocol of LLM-Twin with Universally Composable (UC) \cite{959888} framework based on the design of one-way security and homomorphism in the previous section, which provides a more comprehensive framework for proving the security of LLM-Twin. In addition, we show the key notations and parameters of this section in Table \ref{tab:key_notations}.

\begin{table}[h!]
\centering
\begin{tabularx}{\textwidth}{|l|X|}
\hline
\arrayrulecolor{titlegray}
\hline
\rowcolor{titlegray}
\textbf{Notation} & \textbf{Interpretation} \\
\arrayrulecolor{black}
\hline
$M$ & A decision model built on physical data, which can react based on the rules and states of real devices.\\
\hline
$N$  & Virtual networks of the DTN, which are generally modeled through computation, data access, and so on. \\
\hline  
$H$  & Historical data (including state, rules, etc.) for physical entities used to construct DTs. \\
\hline  
$S$   & Current state information for constructing real-time replicas of physical entities. \\
\hline  
$d$  & Information from other DTs via the virtual network. (inter-twin communication) \\
\hline  
$w$  & Machine Learning Models, training parameters. \\
\hline  
$\varepsilon$   & The number of CPU cycles required to execute a unit of model training. \\
\hline  
$\alpha$  & The number of CPU cycles required to execute a unit of model aggregation. \\
\hline  
$f$  & The CPU cycle frequency. \\
\hline  
$u$  & Physical entities or edge devices corresponding to digital twins. \\
\hline  
$g$  & Cloud servers or edge servers for building digital twins. \\
\hline  
\end{tabularx}
\caption{Key notations used in this paper.}
\label{tab:key_notations}
\end{table}



\subsection*{Computation and communication model}

The traditional FL-based DTN paradigm $DT_i(t)$ \cite{9145588} is shown in Eq.(\ref{eq1}), which contains the decision model $M_i$, the virtual network $N_i$, the historical data $H_i$, the state information $S_i$, and the shared data $d_i$, which are synchronously mapped and constructed from the physical entities $u_i$. Specifically, $DT_i$ continuously interacts with the physical entity $u_i$ to maintain consistency, which involves model synchronization and state synchronization. DT trains the model by collecting historical data and state information, which is used to analyze and make decisions based on the currently synchronized real-time state and shared data. In contrast, we propose LLM-Twin and redesign $\widetilde{DT}_i(t)$ in Eq.(\ref{eq1}). The cloud LLM model $\widetilde{M}_i$ contains the static state information $S_{s,i}$ of the physical entities when synchronizing the model, without additional state synchronization, as shown in Fig. \ref{fig2}. Meanwhile, the $\widetilde{M}_i$ itself is a virtual network where the DTs can share their information, which means that additional construction is not required. For the history information $\widetilde{H}_i$, each physical entity uploads from local to the prompt database via semantic communication, which contains prompt history, real-time state, and shared data.

\begin{flalign}
\begin{aligned}
\left\{\begin{array}{l}
DT_i(t) = \Gamma(M_i, N_i, H_i, S_i, d_i, t)\\
\widetilde{DT}_i(t) = \Gamma(\widetilde{M}_i, \widetilde{H}_i, t) \\
\widetilde{M}_i  = M_i + N_i + S_{s,i} \\
\widetilde{H}_i = \text{Prompt}(u_i,...,u_n) =\sum_{i  = 1}^{n}(H_i+d_i+S_{d,i})
\end{array}\right.
\end{aligned}
\label{eq1}
\end{flalign}

In the traditional FL scheme, the DT model is constructed by distributed training. The physical entity $u_i$ trains local weights using local historical data based on the loss function

\begin{flalign}
\begin{aligned}
&L_i(w) = \frac{1}{\left | H_i \right |} \sum_{x_j,y_j \in H_i} l(w, x_j, y_j)\\
&L_g(w) = \frac{1}{\left | H_g \right |} \sum_{i=1}^{n} L_i(w)
\end{aligned}
\end{flalign}
where $x_j,y_j \in H_i$ is the samples of training data. DT will aggregate the weights of each entity and perform the next round of distributed training with the goal of minimizing the aggregation loss function $L_g(w)$, where $\left | H_g \right | = \sum_{i=1}^{n} \left | H_i \right |$ is the total size of data from participating physical entities. Specifically, the local parameter update and the global parameter aggregation are shown in Eq. (\ref{eq3}).

\begin{equation}
\begin{aligned}
&w_i(t)  = w(t - 1) - \eta \nabla L_i(w(t - 1))\\
& w(t)  = \frac{1}{\left | H_g \right | } \sum_{i = 1}^{n} H_i w_i(t)
\end{aligned}
\label{eq3}
\end{equation}

However, the full-parameter training and multiple rounds of iterations of the above schemes result in significant computation and communication overheads. Therefore, we propose the LLM-Twin with edge fine-tuning based on LoRA \cite{hu2021lora} in Eq.(\ref{eq4}), which only requires training a few parameters to get a good alignment for LLM.

\begin{equation}
\max_{w} \sum_{(x,y) \in Z} \sum_{t=1}^{|y|} \log \left( P_{w + \Delta \widetilde{w}} \left( y_t | x, y_{<t} \right) \right), \quad \left |  \widetilde{w}  \right | \ll \left | w \right |   
\label{eq4}
\end{equation}

The above equation shows the LLM is initialized with weights $w$ and updated with $w + \Delta \widetilde{w}$ by maximizing the conditional language modeling objective, where $\widetilde{w}$ is much smaller than $w$, 1\% or less. Furthermore, based on the above preliminary knowledge, we can obtain the time consumption $T$ of the traditional DTN and the time consumption $\widetilde{T}$ of the proposed LLM-Twin. We define the CPU cycle frequency of the physical entity $u_i$ as $f_{u_i}$, $\xi$ denotes the number of CPU cycles required for each data unit when training the model, and $\alpha$ denotes the number of CPU cycles required for each parameter unit when aggregating the parameters by the DT server. Finally, we get

\begin{equation}
\begin{aligned}
T^{cmp}_{u_i} = \frac{\xi_i \left | H_{i} \right |}{f_{u_i}} \left | w_{i} \right |, \quad  T_{g_j}^{cmp} = \frac{\alpha_j\sum_{i=1}^{n} \left | w_{i} \right | }{f_{g_j}}
\\
\end{aligned}
\label{eq5}
\end{equation}

\begin{equation}
\begin{aligned}
\widetilde{T} ^{cmp}_{u_i} = \frac{\xi_i \left | S_{s,i} \right |}{f_{u_i}} \left | \widetilde{w}_{i}  \right |, \quad  \widetilde{T}_{g_j}^{cmp} = \frac{\alpha_j \left | \widetilde{w}_{i} \right | }{f_{g_j}}, \quad \left | S_{s,i} \right | \ll \left | H_{i} \right|, \left | \widetilde{w}_{i} \right | \ll \left | w_{i} \right | 
\end{aligned}
\end{equation}
where $T^{cmp}_{u_i} $ and $T_{g_j}^{cmp}$ respectively denote the physical entity computation time and the server computation time for constructing a DTN based on the traditional FL; similarly, $\widetilde{T} ^{cmp}_{u_i}$ and $\widetilde{T}_{g_j}^{cmp}$ denote the computation time of LLM-Twin. In addition, to further model the communication consumption, we define the data transmission rate between a physical entity $u_i$ and a DT server $g_j$

\begin{equation}
\begin{aligned}
&r_{u_i,g_j} = \frac{c_{u_i}}{C_0}B\log{(1+\gamma_{u_i,g_j})} \\
&\sum_{i=1}^{n} c_{u_i} + \sum_{j=1}^{m} c_{g_j} \leq C_0
\end{aligned}
\end{equation}
where $C_{0}$ denotes the total number of subchannels over the whole bandwidth $W$, the number of subchannels allocated to the physical entity $u_i$ is $c_{u_i}$, and $\gamma_{u_i,g_j}$ denotes the channel state. Further, we define the intra-twin communication time $T^{intra}_{u_i}$ and the inter-twin comminication time $T^{inter}_{g_j}$ for the FL method in Eq.\ref{eq8}, where $T^{intra}_{u_i}$ contains the data size $\left | w_i(t) \right |$ for model synchronization and the data size $\left | S_i \right |$ for entity state synchronization. And then, $T^{inter}_{g_j}$ is modeled as the time required for parameter aggregation in DT server.

\begin{equation}
\begin{aligned}
T^{intra}_{u_i} = \frac{\left | w_i(t) \right | + \left | S_i \right |}{r_{i}}, \quad T^{inter}_{g_j} = T^{cmp}_{g_j}
\end{aligned}
\label{eq8}
\end{equation}

Similarly, we get the LLM-Twin communication time

\begin{equation}
\begin{aligned}
&\widetilde{T} ^{intra}_{u_i}  = \frac{\left | \widetilde{w}_i (t) \right | + \left | \widetilde{S}_{d,i} \right |}{r_{i}}, \quad \widetilde{T} ^{inter}_{g_j}  = \frac{\left | \widetilde{H}_i  \right | }{r_{N_j}} \\
&\widetilde{S}_{d,i} = Semantic(S_{d,i}), \quad \left | \widetilde{S}_{d,i} \right |  \ll \left | S_{d,i} \right | \ll \left | S_i \right | , \quad r_{N_j} \gg r_j
\end{aligned}
\label{eq9}
\end{equation}
where $\left | \widetilde{S}_{d, i} \right |$ denotes the state information that needs to be transmitted dynamically, and $r_{N_j}$ denotes the transmission rate of the virtual network, such as the parameter aggregation computation rate of the server in the FL method. Obviously, DTNs rely on communication and computation for their construction and maintenance, so communication and computation are the core components for analyzing the efficiency of DTNs. Therefore, we obtain the time consumption $T_{fl}$ of the FL-based DTN construction method by counting the total communication time and computation time in Eq. (\ref{eq10}).

\begin{equation}
\begin{aligned}
T_{fl} &= K (\max \left \{ T^{cmp}_{u_i}, i=1,...,n \right \} + T^{cmp}_{g_i} + \frac{\left |  w_i(t)\right | + \left | w_i(t+1) \right |  }{r_i} ) + \frac{\left | H_i(t) \right | }{r_i}\\
&= K(\frac{\xi_i \left | H_i(t) \right |  \left | w_i(t) \right | }{f_{u_l}} + \frac{\alpha_j \sum_{i=1}^{n}{\left | w_i(t) \right | }} {f_{g_j}}) + \frac{K(\left | w_i(t) \right | +\left | w_i(t+1) \right | ) + \left | H_i(t) \right | }{r_i} 
\end{aligned}
\label{eq10}
\end{equation}

Similarly, we obtain the time efficiency equation for LLM-Twin

\begin{equation}
\begin{aligned}
T_{LLM-Twin} & = \frac{1}{\lambda}(\widetilde{T} ^{cmp}_{u_i} + \widetilde{T} ^{cmp}_{g_j} + \frac{\left | \widetilde{w} _i(t) \right | }{r_i} ) + \frac{\left | \widetilde{S} _{d,i} \right | }{r_i} + \widetilde{T} ^{inter}_{s_j}   \\  
& = \frac{1}{\lambda}(\frac{\xi_i \left | S_{s,i} \right | }{f_{u_i}} \left | \widetilde{w} _i(t) \right |   + \frac{\alpha_j \left | \widetilde{w} _i(t) \right |  }{f_{g_j}} + \frac{\left | \widetilde{w} _i(t) \right | }{r_i} ) + \frac{\left | \widetilde{S} _{d,i} \right | }{r_i} + \frac{\left | \widetilde{H} _i(t) \right | }{r_{N_j}} 
\end{aligned}
\label{eq11}
\end{equation}
where $K$ denotes the number of rounds required for convergence since FL involves multiple iterations. Differently, as mentioned in the previous section, LLM-Twin is not required to update the static information (semantic knowledge base) in real-time, so $\lambda$ denotes the periodicity of the update of the static information. Finally, according to the following constraints

\begin{flalign}
\begin{aligned}
\left\{\begin{array}{l}
\left | \widetilde{S} _{d,i} \right | \ll \left | S_{i} \right |,\quad \left | S_{s,i} \right | < \left | S_{i} \right |, \quad \left | \widetilde{w}_i \right | \ll \left | w_i \right | \\
\left | S_{s,i} \right | \left | \widetilde{w} _{i} \right |  -   \left | S_{i} \right |\left | w_{i} \right | < 0\\
\left | \widetilde{w} _{i} \right | - \sum_{i=1}^{n}\left | w_{i} \right | < 0\\
(\left | \widetilde{w} _{i} \right | + \left | \widetilde{S} _{d,i} \right |) -  (\left | w_{i} \right | + \left | S_{i} \right |) < 0\\
\end{array}\right.
\end{aligned}
\end{flalign}
we will compare the efficiency of LLM-Twin with the traditional FL method. To demonstrate the advantages of LLM-Twin more intuitively, we assume that the FL method completes in just one round of iterations, i.e., $K=1$. In addition, it is assumed that LLM-Twin synchronizes static information in real-time, i.e., $\lambda$ = 1. Despite this assumption, we still get that LLM-Twin consumes less time than the FL approach in Eq.(\ref{eq13}).

\begin{flalign}
\begin{aligned}
T_{LLM-Twin} - T_{fl} &= (\frac{\xi_i \left | \widetilde{S} _{s,i} \right | \left | \widetilde{w} _i(t) \right | - \xi_i \left | S_i \right |  \left | w_i (t) \right |   }{f_{u_i}}) 
+ (\frac{\alpha_j \left | \widetilde{w}_i(t) \right | - \alpha_j \sum_{i = 1}^{n} \left | w_i(t) \right | }{f_{g_j}} )\\
&+ \frac{(\left | \widetilde{w} _i(t) \right | + \left | \widetilde{S} _{d,i} \right |) - (\left | w_i(t) \right | + \left | S_i \right | ) }{r_i} + \frac{\left | \widetilde{H} _i(t) \right | }{r_{N_j}} \quad < \quad 0 
\end{aligned}
\label{eq13}
\end{flalign}

From the above mathematical analysis, it can be seen that LLM-Twin is efficient mainly in the following aspects: 1) Instead of synchronizing all state information $S_i$ in real time, LLM-Twin only needs to synchronize the dynamic semantic information $\widetilde{S} _{d,i}$ in it; 2) LLM-Twin does not need to perform parameter aggregation and iteration of participating entities to accomplish data sharing while training the model, instead, it accomplishes data sharing during model inference by searching the prompt database $\widetilde{H} _i(t)$; 3) Instead of training the model with full parameters $w_i(t)$, LLM-Twin only needs to fine-tune a very few parameters $\widetilde{w} _i(t)$.

\subsection*{Security analysis of LLM-Twin}

In this section, we formalize LLM-Twin as a service protocol
\begin{table}[h!]
\centering
\begin{tabularx}{\linewidth}{|X|}
\arrayrulecolor{black}
\hline
\rowcolor{titlegray}
\multicolumn{1}{|l|}{\textbf{Protocol Description}} \\
\arrayrulecolor{black}
\hline
\begin{description}[leftmargin=!, font=\normalfont\bfseries\itshape, topsep=1pt, partopsep=0pt, parsep=0pt, itemsep=1pt]
  \item[Data Upload:]\mbox{}
  \begin{enumerate}[nosep]
    \item Data provider A possesses data $(E, P)$.
    \item A performs some local operation to transform $(E, P)$ to $(E', P')$.
    \item A uploads $(E', P')$ to third-party C.
  \end{enumerate}
  
  \item[Service Function Computation:]\mbox{}
  \begin{enumerate}[nosep]
    \item Upon receiving $(E', P')$, C processes it to obtain $F(E, P)$.
  \end{enumerate}

  \item[Data Retrieval:]\mbox{}
  \begin{enumerate}[nosep]
    \item Service requestor B sends a request $q$ to C.
    \item C computes $L(q, F(E, P))$, retrieves $P$, and sends it to B.
  \end{enumerate}
\end{description}
\\ \hline
\end{tabularx}
\end{table}
 ,where $(E, P)$ denotes the training data pair, $E$ is the LLM prompt, and $P$ is the corresponding completion. $(E,P)$ is fine-tuned to $(E',P')$ by edges and uploaded to C. C denotes the DT server that obtains the DT model $F(E, P)$ by merging the fine-tuning parameters, which can respond to B's request q, and then respond with the completion $P$ to B. Further, the protocol has the following properties

\begin{table}[h!]
\centering
\begin{tabularx}{\linewidth}{|X|}
\arrayrulecolor{black}
\hline
\multicolumn{1}{|l|}{\cellcolor{titlegray}\textbf{Security Properties}} \\
\hline
\begin{enumerate}[noitemsep]
  \item $F(E, P)$ can be used to obtain $P$ if the $q$ corresponding to $E$ is known. 
  \item Knowing $P$ alone, one cannot retrieve $E$.
  \item $F(E, P)$ and $(E', P')$ do not allow the recovery of $E$ or $P$, ensuring the privacy of the original data.
\end{enumerate}
\\ \hline
\end{tabularx}
\end{table}
It is obvious by the principle of LLM that $q$ as a prompt can obtain the corresponding completion $P$. From the one-way security in the previous section, as shown in Fig. \ref{fig4}, it is not possible to derive the training-time sensitive data $E$ by completion $P$, and It is not possible to recover the original $E$ and $P$ from the model and weights. Similarly according to the homomorphic design, as in Fig. \ref{fig3}, the LLM model and the fine-tuned weights can accomplish the service while guaranteeing $E$ and $P$ privacy. Based on the above protocol and security properties, we define the ideal-world LLM-Twin function $\mathcal{F}_{DATA}$.
\begin{table}[h!]
\centering
\begin{tabularx}{\linewidth}{|X|}
\arrayrulecolor{black}
\hline
\multicolumn{1}{|l|}{\cellcolor{titlegray}\textbf{Ideal Functionality $\mathcal{F}_{DATA}$}} \\
\hline
\begin{description}[leftmargin=!, font=\normalfont\bfseries\itshape, topsep=1pt, partopsep=0pt, parsep=0pt, itemsep=1pt]
  \item[Upon receiving $("\text{upload}", A, C, E, P)$ from A:]\mbox{}
  \begin{enumerate}[nosep]
    \item Store $(E, P)$ internally.
    \item Send $("\text{receipt}", A, C)$ to C.
  \end{enumerate}
  \item[Upon receiving $("\text{request}", B, C, q)$ from B:]\mbox{}
  \begin{enumerate}[nosep]
    \item Compute $P$ using $E$ and $F(E, P)$.
    \item Send $("\text{response}", B, C, P)$ to B.
  \end{enumerate}
  \item[If B is corrupted:]\mbox{}
  \begin{enumerate}[nosep]
    \item Upon receiving $("\text{corrupt}", B)$ from $\mathcal{S}$, mark B as corrupted.
    \item If simulator $\mathcal{S}$ issues a $("\text{request}", B, C, q)$, simulate $P$ or altered $P'$ (if $\mathcal{S}$ wants to simulate a cheating B) based on $E$ and $F(E, P)$ and send $("\text{response}", B, C, P')$ to B.
  \end{enumerate}
\end{description}
\\ \hline
\end{tabularx}
\end{table}

In ideal functionality $\mathcal{F}_{DATA}$, A refers to the physical entity that performs edge fine-tuning, C refers to the DT model, and B refers to the service requester, e.g., other DTs. The important point is that $\mathcal{F}_{DATA}$ is security as it is based on a fully trusted and ideal third party. In addition, external environments$ \mathcal{Z}$ and adversaries $\mathcal{A}$ will observe and attack the protocol. As a result, the simulator $\mathcal{S}$ is used to simulate the attack behavior of the adversary $\mathcal{A}$ in the ideal world and interacts with the external environment $\mathcal{Z}$, making it impossible for $\mathcal{Z}$ to distinguish between the ideal-world protocols and the real LLM-Twin protocol.

\begin{table}[h!]
\centering
\begin{tabularx}{\linewidth}{|X|}
\arrayrulecolor{black}
\hline
\multicolumn{1}{|l|}{\cellcolor{titlegray}\textbf{Simulator}} \\
\hline
\begin{description}[leftmargin=!, font=\normalfont\bfseries\itshape, topsep=1pt, partopsep=0pt, parsep=0pt, itemsep=1pt]
  \item[Simulator $\mathcal{S}$ when corrupting A:]\mbox{}
  \begin{enumerate}[nosep]
      \item A's corruption isn't significantly impactful because the transformation from $(E, P)$ to $(E', P')$ occurs before the data's interaction with C. The simulator simply follows the honest protocol on behalf of A.
  \end{enumerate}
  
  \item[Simulator $\mathcal{S}$ when corrupting C:]\mbox{}
  \begin{enumerate}[nosep]
      \item Upon receiving $(E', P')$ from adversary $\mathcal{A}$:
  \begin{itemize}[noitemsep]
    \item Send $("\text{upload}", A, C, \text{dummy\_E}, \text{dummy\_P})$ to $\mathcal{F}_{DATA}$.
  \end{itemize}

  \item When $\mathcal{A}$ processes a request from B:
  \begin{itemize}[noitemsep]
    \item Send $("\text{request}", B, C, q)$ to $\mathcal{F}_{DATA}$.
    \item Upon receiving $P$ from $\mathcal{F}_{DATA}$, send it to $\mathcal{A}$.
  \end{itemize}
  \end{enumerate}
  \item[Simulator $\mathcal{S}$ when corrupting B:]\mbox{}
  \begin{enumerate}[nosep]
  \item Upon B's corruption, $\mathcal{S}$ sends $("\text{corrupt}", B)$ to $\mathcal{F}_{DATA}$.
  \item $\mathcal{S}$ intercepts $q$ from $\mathcal{A}$.
  \item $\mathcal{S}$ sends $("\text{request}", B, C, q)$ to $\mathcal{F}_{DATA}$.
  \item When $\mathcal{F}_{DATA}$ responds with $P$, $\mathcal{S}$ relays this to $\mathcal{A}$ or alters the message to $P'$ if $\mathcal{S}$ simulates a cheating B trying to alter the data.
  \end{enumerate}
\end{description}
\\ \hline
\end{tabularx}
\end{table}
Based on the above simulator design, for every adversary $\mathcal{A}$ in the real world, there exists an ideal-world simulator $\mathcal{S}$ and an ideal function $\mathcal{F}_{DATA}$ such that there is no environment $\mathcal{Z}$ that can distinguish between the two worlds. Specifically, 1) Neither $\mathcal{A}$ in the real world nor $\mathcal{S}$ in the ideal world can extract original training data $E$ from interactions involving only $P$. 2) $\mathcal{S}$ can simulate any actions by altering the response to $\mathcal{A}$ or by sending modified queries. The ideal world captures any dishonest behavior that $B$ could exhibit in the real world. 3) Given the simulator's capabilities and the ideal functionality, any environment $\mathcal{Z}$ cannot distinguish between the real world. Therefore, the LLM-Twin security framework is given based on UC. Since the ideal world is security and the ideal world is made indistinguishable from the real world by constructing ideal functions and simulators, it is demonstrated that the real-world protocol, i.e., LLM-Twin, is UC security.

\section*{Evaluation and case study}
In this section, we first show the numerical experimental results of LLM-Twin in terms of computation and communication consumption and demonstrate the advantages of LLM-Twin by comparing its performance with traditional FL-based DTNs. Then, we show a case study of a smart home digital twin network, further demonstrating the LLM-Twin's feasibility and benefits. All of the experiments are deployed on the high-performance server with the configuration of Intel Xeon Gold 6226R
3.90 GHz CPU, 256 GB RAM, Ubuntu 20.04 LTS and Python 3.8. 

\subsection*{Performance analysis}
\begin{figure}[h!]
\includegraphics[width=11cm]{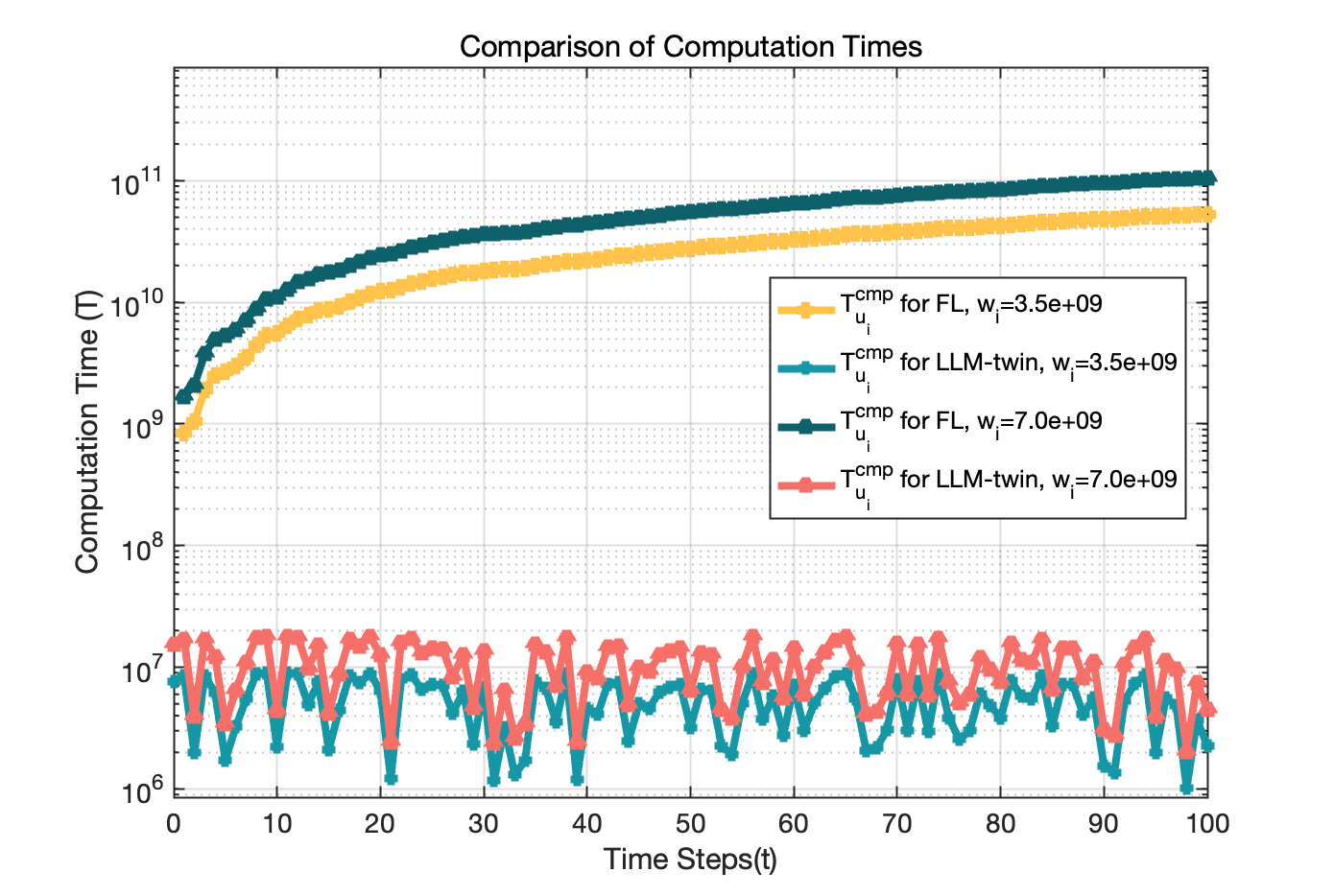}
\caption{Comparison of physical entity computation consumption between LLM-Twin and FL-based DTN.}
\label{fig5}
\end{figure}

In the FL method, the edge physical entities use the historical data to train the model locally and get the distributed model weights, where the time consumed for this part is denoted as $T^{cmp}_{u_i}$. Moreover, the time when the weights of all physical entities are aggregated in the DT server is denoted as $T^{cmp}_{g_i}$. In LLM-Twin, the physical entities use current static state data to fine-tune locally for obtaining the small LM, and this part of the computation time is denoted as $\widetilde{T}^{cmp}_{u_i}$. The weight files of the small model are loaded into the large model in the DT server, and this part of the time is denoted as $\widetilde{T}^{cmp}_{g_i}$. Subsequently, we evaluated the computational performance of the FL method and LLM-Twin respectively in two cases with total parameters $w$ of 3.5B and 7B, as shown in Fig. \ref{fig5}.

\begin{figure}[h!]
\includegraphics[width=11cm]{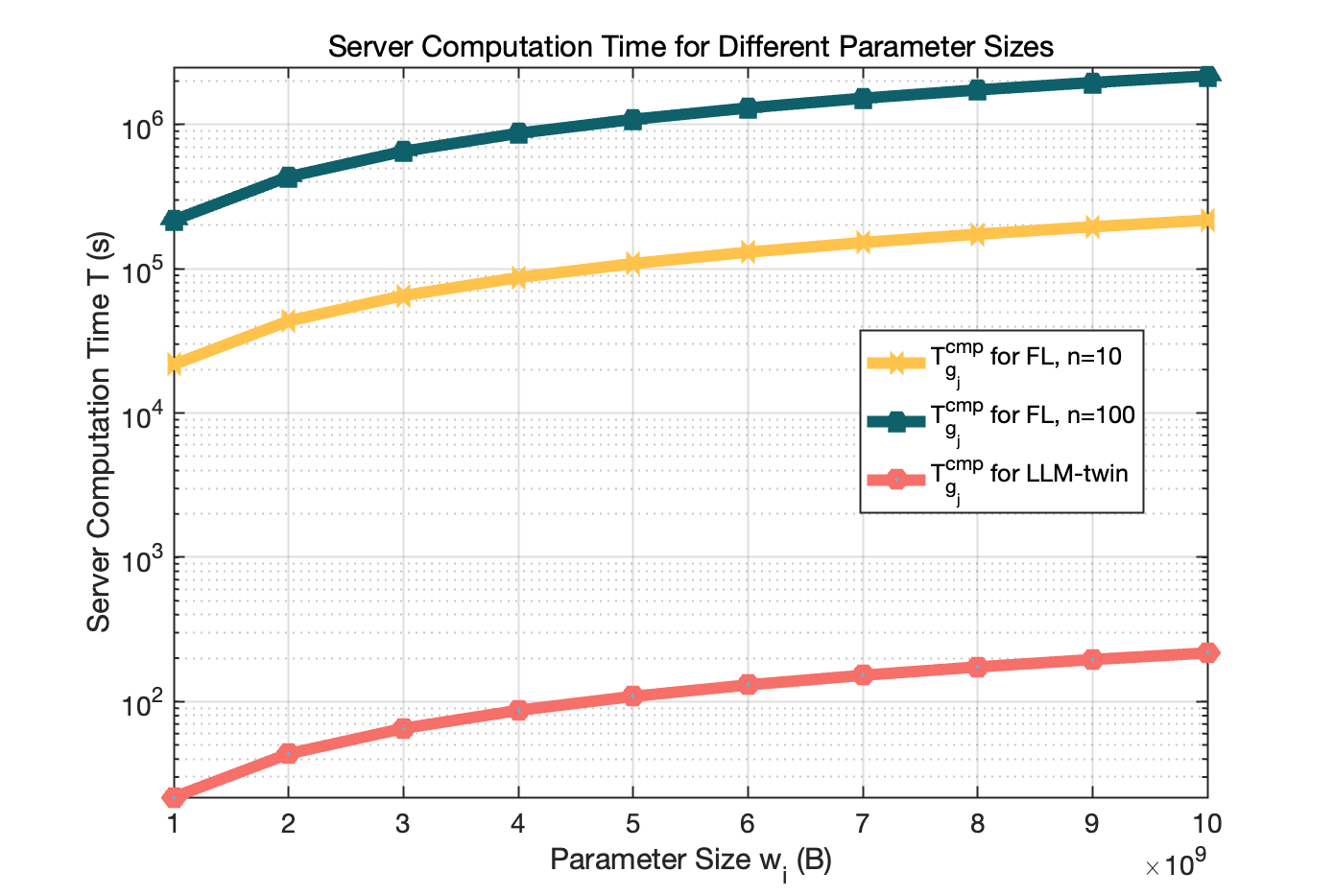}
\caption{Comparison of DT server computation consumption between LLM-Twin and FL-based DTN.}
\label{fig6}
\end{figure}

As shown above, LLM-Twin consumes only \%1 of FL or even less time in edge physical entity training. Moreover, the time consumed by the FL method will continue to increase over time as more historical data becomes available. Further, Fig. \ref{fig6} illustrates the time required for the DT server to aggregate the weights as the parameter size increases for different numbers of physical entities. Since FL needs to aggregate the weights of each entity in the computation phase to accomplish information sharing, the number of entities will significantly affect $T^{cmp}_{g_i}$. In contrast, in the computation phase, LLM-Twin does not need to consider the parameters of the other DTs but only loads the weights of the corresponding entities. Therefore $\widetilde{T}^{cmp}_{g_i}$ will not be affected by the number of entities and also less computation leads to much less computation time than the FL method, as shown in Fig. \ref{fig6}.

\begin{figure}[h!]
\includegraphics[width=11cm]{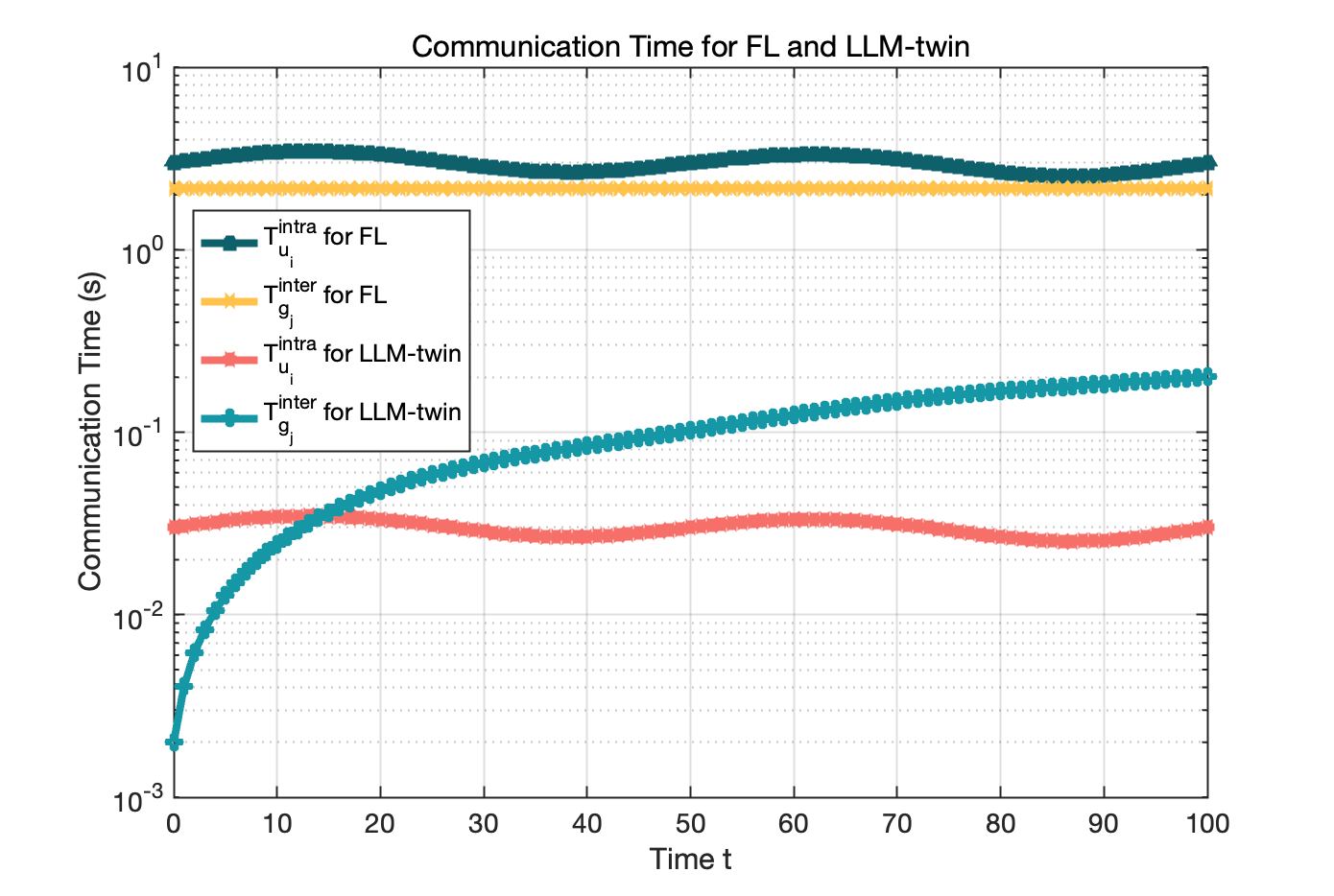}
\caption{Comparison of communication consumption between LLM-Twin and FL-based DTN.}
\label{fig7}
\end{figure}

Next, the communication performance will be evaluated, including the core communication architectures of the DTN: intra-twin communication and inter-twin communication. In the FL approach, intra-twin communication is denoted as $T^{intra}_{u_i}$, which includes the model's weight update and the state synchronization of the physical entities. In FL, inter-twin communication is modeled as the weight aggregation of each entity, denoted as $T^{inter}_{g_i}$, which is the same as $T^{cmp}_{g_i}$, as shown in Eq. \ref{eq8}. In contrast, intra-twin communication for LLM-Twin is modeled as the synchronization of the semantic knowledge base, denoted as $\widetilde{T}^{intra}_{u_i}$, which is achieved by transmitting the fine-tuned weights to the DT server. Inter-twin communication is modeled as searching context information from the prompt database, denoted as $\widetilde{T}^{inter}_{g_i}$. Finally, the results of the evaluation are shown in Fig \ref{fig7}.

\begin{figure}[h!]
\includegraphics[width=11cm]{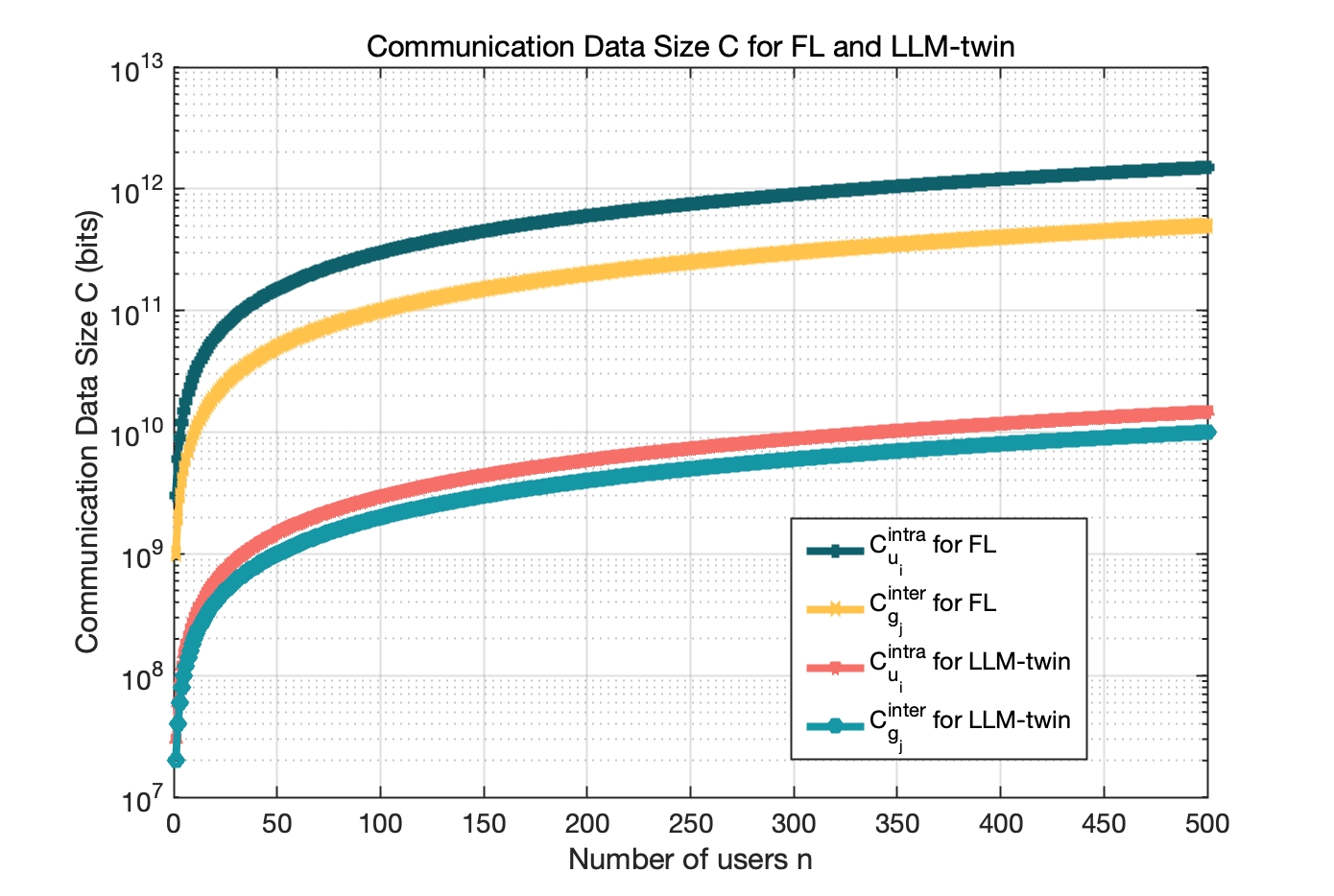}
\caption{Comparison of communication content size between LLM-Twin and FL-based DTN.}
\label{fig8}
\end{figure}

Since the communication content of FL is always model parameters and state information, the communication cost will not be affected by time. The inter-twin communication of LLM-Twin requires searching for a specific context in the prompt database. Since the prompt database continuously accepts dynamic information from all physical entities and stores it, it generates a large number of historical records as time advances and thus impacts $\widetilde{T}^{inter}_{g_i}$. Most importantly, the LLM-Twin is based on a fine-tuned mini-giant modeling scheme that allows far fewer parameters to be transmitted in the communication than the FL approach. In addition, LLM-Twin's semantic communication-based approach makes intra-twin communication transmit less state information data, as shown in Eq. \ref{eq9}. Therefore, the communication efficiency of the LLM-Twin is still much higher than the FL, as shown in Fig \ref{fig7}, and the communication content is much less than the FL, as shown in Fig \ref{fig8}.

\begin{figure}[h!]
\includegraphics[width=11cm]{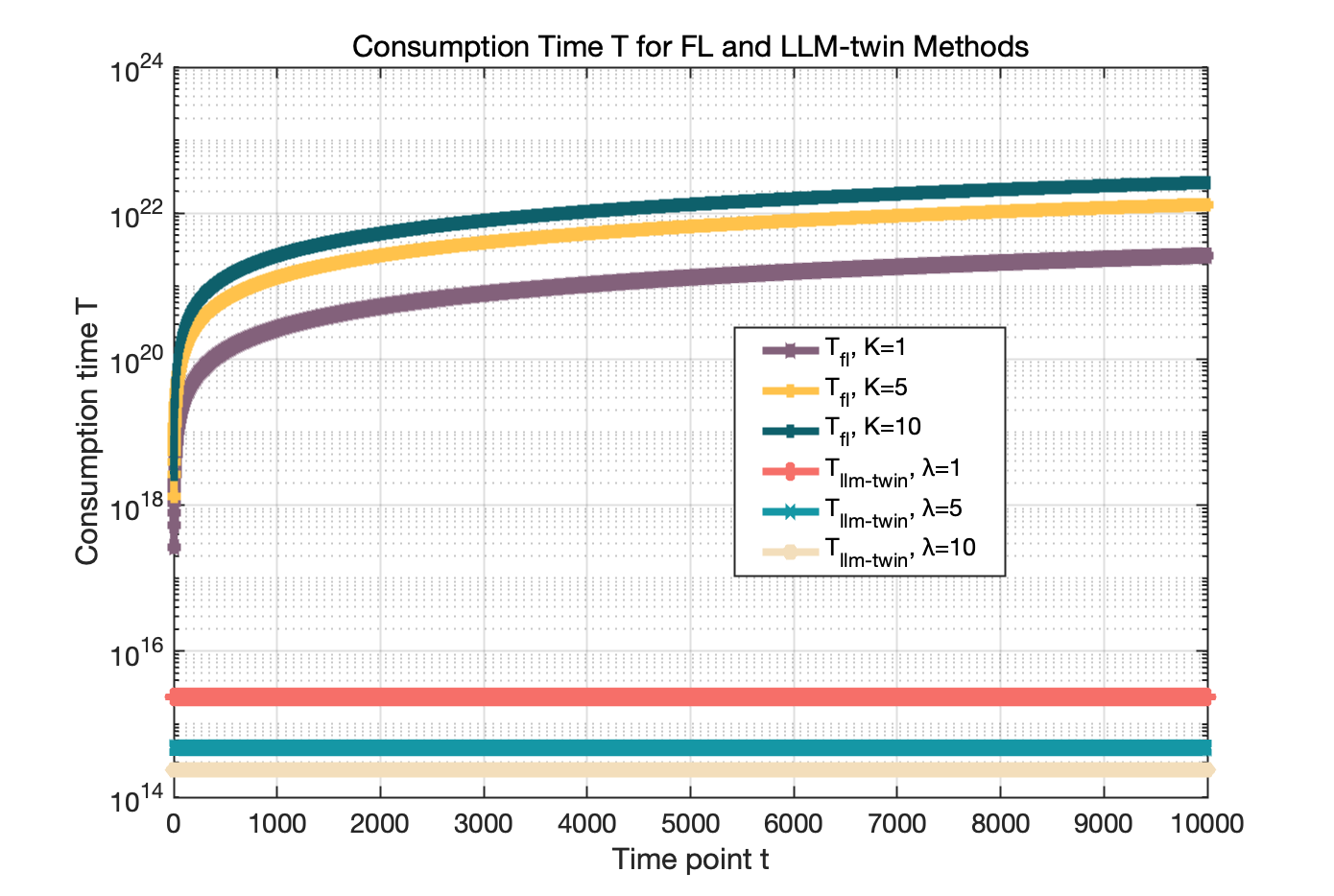}
\caption{Comparison of overall DTN consumption between LLM-Twin and FL-based DTN.}
\label{fig9}
\end{figure}

Finally, we compare the overall efficiency of DTNs constructed based on FL and LLM-Twin, and the evaluation results are shown in Fig. \ref{fig9}. Since FL requires multiple rounds of weight updates to complete model convergence, which means that the number of iteration rounds significantly affects the DTN efficiency. On the contrary, LLM-Twin does not need to update the entity static information and model in real-time, as shown in Eq. \ref{eq11}, which brings higher efficiency. In addition, it can be seen from Fig. \ref{fig9} that the accumulation of historical data due to time passing has a significant impact on the efficiency of FL-based DTN, but has less impact on LLM-Twin.

\subsection*{Case study in smart home digital twin network}

\begin{figure}[h!]
\includegraphics[width=16cm]{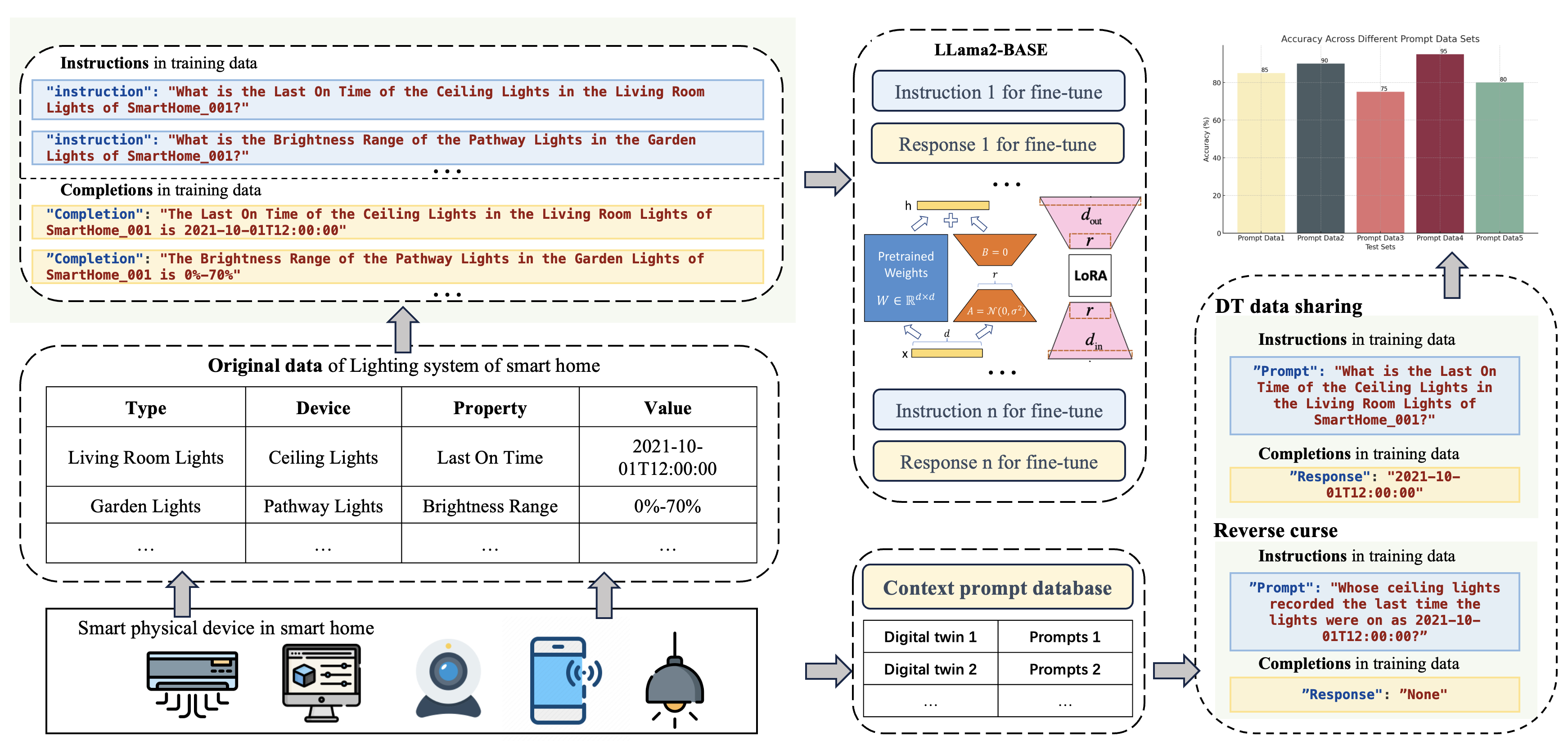}
\caption{Case study of LLM-Twin in smart home digital twin network.}
\label{fig10}
\end{figure}
\begin{figure}[h!]
\includegraphics[width=11cm]{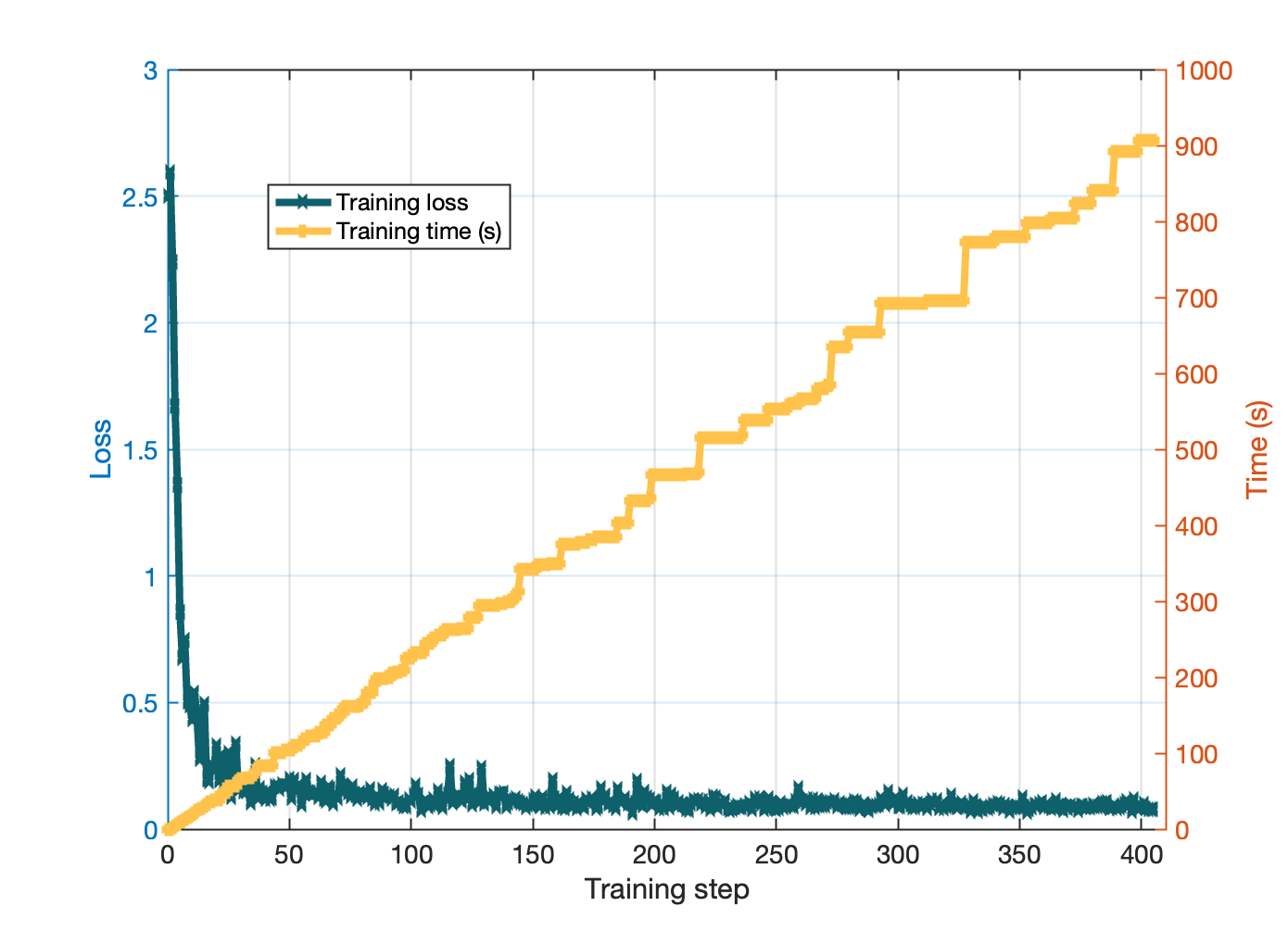}
\caption{Training losses and training time.}
\label{fig11}
\end{figure}

In order to further demonstrate the feasibility and practical performance of LLM-Twin, the case study based on a smart home digital twin network is designed in this section, as shown in Fig. \ref{fig10}. We first collected and generated 1.6k state data of the smart home, which includes system types, devices, property, and state values. This data will be encoded into an LLM fine-tuned dataset in a local Python environment, which will follow the form <prompt, completion>. Then we execute Lora fine-tuning with the LLama7B model locally. The proposed mini-giant framework demonstrates its advantages. The memory footprint during local fine-tuning is 5878 MB, which satisfies the performance requirements of most edge devices. In addition, the loss and training time for model fine-tuning are shown in Fig. \ref{fig11}. The model basically converges after 100 seconds of training and regionally stabilizes after about 400 seconds. As mentioned previously, LLM-Twin updates the static data at regular intervals, so the latency of approximately 400 seconds for local fine-tuning is optimistic. After completing the fine-tuning, we uploaded the fine-tuned weights file to the server and loaded the local information into LLama7B through weight merging. The weight merging consumes about 19820 MB of memory, which is acceptable for a DT server. Further, we upload the prompt which describes the dynamic information of the physical entity to the server's database, and the intra-twin communication of LLM-Twin has been completed. 

In inter-twin communication, the DT owner constructs the prompt to ask the LLM for information based on actual demands. The LLM loads the local weights of the DT owner and then generates a response by combining the context in the prompt database. For example, $smarthome1$ wants to adjust the brightness of its lights to make it higher than the average of all the homes in the region. the LLM of $DT1$ accepts the request and fetches prompts from other DTs that contain the brightness data from the prompt database. These prompts will add to the context and combined with the original request to be used as the final prompt. Since LLM of $DT1$ has already loaded static information about $smarthome1$ via intra-twin communication, it can make decisions based on the properties of $smarthome1$. For example, if $smarthome1'$s light brightness is below average $A$, but the brightness range is above average $A$, it will tell $DT1$ to increase the light brightness to $A$. 

\begin{table}[h!]
\setlength{\arrayrulewidth}{0.5pt}
\arrayrulecolor{black}
\begin{tabular}{|c|c|c|c|}
\hline
\rowcolor{titlegray} 
Prompt type& Prompt number& Correct responses number& Incorrect responses number\\
\hline
Normal & 100 & 94 & 6\\
\hline
Normal & 500 & 485 & 15\\
\hline
Normal & 1000 & 912 & 88\\
\hline
Reverse & 500 & 0 & 500\\
\hline
Reverse & 1000 & 0 & 1000\\
\hline
\end{tabular}
\caption{Accuracy test of LLM-Twin's information generation.}
\label{table2}
\end{table}

For inter-twin communication, the time for executing queries and computations inside the server is negligible, and we focus on the accuracy of inter-twin communication. We generated a large number of test prompts to detect whether LLM can correctly generate information about its DTs. Table \ref{table2} shows the final results, where LLM can generate responses with 90\% accuracy in all tests. Therefore the proposed LLM-Twin learns the state information of the physical entities, which contributes to further decision-making based on the state information of the physical entities.

In addition, we further validate the proposed data security design. When training LLM-Twin, we strengthen the reversal curse of LLM by fixing the training set's order of discourse, e.g., $<object, description>$. $object$ is known only to the DT owner. Therefore, an attacker cannot obtain the DT owner's $object$ by submitting a large number of $descriptions$ to LLM. For example, an attacker submits "Who has lights brighter than 80\%?" , LLM will reply with garbled code. Table \ref{table2} shows the results of our test, where the accuracy of the reversed prompt is 0. This verifies the one-way security of LLM-Twin.

\section*{Conclusion}

Beyond 5G networks bring huge benefits to efficient and secure IoT, IoE, etc., especially where digital twin networks make it possible to efficiently process massive amounts of data and make real-time optimizations. Traditional DTNs networking frameworks face difficulties in dealing with massive amounts of multi-type data and ensuring efficient communication and computation. LLM brings opportunities for DTNs due to its powerful multimodal data processing capabilities and strong intelligence. In this paper, We first proposed LLM-based DTN, LLM-Twin. First, the resource-constrained problem faced by LLM deployment in DTN is solved by designing a framework with mini-giant model collaboration. Second, we first proposed the digital twin Semantic network (DTSN), where new intra-twin communication and inter-twin communication designs for DTNs are designed. Numerical experiments demonstrated that DTSN is over 100 times more efficient compared to traditional FL-based DTNs through semantic-level communication and computation. In addition, we consider the data security of LLM-Twin. We design a security model for LLM-Twin and provide a UC security proof for LLM-Twin. Meanwhile, in the use case study we show that by reinforcing the reversal curse, LLM-Twin achieves 100\% defense against reverse attacks. Also in the case study, we verified the feasibility and validity of LLM-Twin, as LLM-Twin demonstrated more than 90\% response accuracy.

\section*{Data availability}
The datasets used and/or analyzed during the current study are available from the corresponding author upon reasonable request.
\bibliography{ref}



\section*{Acknowledgements}

 This work was supported in part by the JSPS KAKENHI under Grants 23K11072, in part by the National Natural Science Foundation of China under Grants U21B2019 and 61972255, and in part by the China Scholarship Council under Grants 202308050055.

\section*{Author contributions statement}

Y.H. and J.W. conceived the idea and proposed the methodology, Y.H. verified the method, Y.H., J.W., and R.M. analyzed the results. All authors reviewed the manuscript.

\section*{Additional information}
\noindent \textbf{Competing interests statement.} The authors declare no competing interests.
\vspace{0.5em}

\noindent \textbf{Correspondence} and requests for materials should be addressed to J.W. 
\vspace{0.5em}

\section*{Legends}
\setcounter{figure}{0} 
\captionof{figure}{[Beyond 5G digital twin networks.]}
\captionof{figure}{[Digital twin semantic network of LLM-Twin networking framework.]}
\captionof{figure}{[Homomorphism model of LLM-Twin to protect original data.]}
\captionof{figure}{[One-way security model of LLM-Twin.]}
\captionof{figure}{[Comparison of physical entity computation consumption between LLM-Twin and FL-based DTN.]}
\captionof{figure}{[Comparison of DT server computation consumption between LLM-Twin and FL-based DTN.]}
\captionof{figure}{[Comparison of communication consumption between LLM-Twin and FL-based DTN.]}
\captionof{figure}{[Comparison of communication content size between LLM-Twin and FL-based DTN.]}
\captionof{figure}{[Comparison of overall DTN consumption between LLM-Twin and FL-based DTN.]}
\captionof{figure}{[Case study of LLM-Twin in smart home digital twin network.]}
\captionof{figure}{[Training losses and training time.]}
\setcounter{table}{0} 
\captionof{table}{[Key notations used in this paper.]}
\captionof{table}{[Accuracy test of LLM-Twin’s information generation.]}

\end{document}